\documentclass[final,5p,authoryear,times]{elsarticle}
\usepackage{graphicx}
\usepackage{subcaption}

\usepackage{amsmath}
\usepackage{xcolor}
\usepackage{mathtools}
\usepackage{stfloats}
\usepackage{booktabs}
\usepackage[nopatch]{microtype}
\newcommand{\rthis}[1]{\textcolor{black}{#1}}
\usepackage{booktabs}
\usepackage[plainpages=false, colorlinks=true, anchorcolor=blue, linkcolor=blue, citecolor=blue, bookmarks=false]{hyperref}
\journal{Astronomy \& Computing}
\begin{document} \sloppy
\begin{frontmatter}

\title{Galaxy Morphology Classification: Uncertainty Modeling and Out of Distribution Detection}

\author[1]{Prem Prakash}\ead{id22resch11013@iith.ac.in}
\address[1]{Center for Interdisciplinary Programs, IIT Hyderabad, Kandi, 502284, Telangana, India}
\author[2]{Shantanu Desai}\ead{shantanud@phy.iith.ac.in}
\address[2]{Department of Physics, IIT Hyderabad, Kandi, 502284, Telangana, India}

\author[3]{P. K. Srijith}\ead{srijith@cse.iith.ac.in}
\address[3]{Department of Computer Science and Engineering, IIT Hyderabad, Kandi, 502284, Telangana, India}

\begin{abstract}
We present a comprehensive framework for galaxy morphology classification that combines  enhanced ``out-of-distribution (OOD)'' detection with improved uncertainty quantification. Using ``Galaxy Zoo DECaLS'', we trained a ResNet-34 architecture under three configurations: standard cross-entropy loss as a baseline, IsoMaxPlus (Isotropy Maximization Plus) loss function for OOD detection, and hybrid IsoMaxPlus+Monte Carlo (MC) Dropout for enhanced uncertainty quantification. IsoMaxPlus replaces the conventional SoftMax logits with distance-based class representations, preserving inter-class separability and enabling reliable OOD detection, without requiring additional architectural modifications or hyperparameter tuning. Coalescing IsoMaxPlus with MC Dropout provides a fast Bayesian approximation by performing multiple stochastic forward passes during inference. Our results show that IsoMaxPlus substantially improves OOD detection, increasing TNR@TPR95 by nearly $90\%$ relative to the cross-entropy baseline, while maintaining a competitive accuracy of 97\% across nine morphological classes. Additionally, with MC Dropout, the model yields more stable predictions and a reduction in calibration error, providing more reliable uncertainty estimates. The Expected Calibration Error (ECE) is reduced from $0.0095$ to $0.0026$ ($\approx73\%$) for IsoMaxPlus and $0.0033$ ($\approx65\%$) when combined with MC Dropout, compared to the baseline. Misclassified or underconfident predictions exhibit higher predictive entropy and lower minimum distance scores, providing an interpretable metric for identifying unreliable predictions. IsoMaxPlus does not require additional training or outlier examples during the training process to produce effective separation between ``in-distribution (ID)'' and OOD samples. These methods are well-suited for current and upcoming large-scale surveys, where reliable automated morphological classification and awareness of uncertainty are essential for identifying rare or previously unseen galaxy morphologies. 
\end{abstract}

\end{frontmatter}

\section{Introduction}

Galaxy morphology has long been a cornerstone of studies of galaxy formation and evolution. 
Galaxy morphology is closely linked to numerous areas of astrophysics and cosmology. Morphological information has been used to investigate stellar masses~\citep{Bundy}, star formation history~\citep{1998ARA&A..36..189K}, investigating  galaxy color, gas, and dust content~\citep{Lianou}, galaxy age~\citep{Bernardi}, studies of dynamical evolution~\citep{Fall}, and even tests of modified gravity theories~\citep{Pedro}. A more comprehensive review of the connection of galaxy morphology to different areas of astrophysics can be found in ~\citet{Conselice14}.

In the early 1920s, ~\citet{Hubble26} visually categorized galaxies into elliptical and spiral based on their shapes and features, introducing the tuning fork model in his revised paper~\citep{Hubble36}. Later, these galaxy categories were refined to include additional features such as lenticular galaxies, ring structures, etc. ~\citep{DeVauc59,Sandage75,Elmegreen87}. More details on the historical developments in galaxy morphology classification can be found elsewhere~\citep{Abraham01,Masters}.
However, morphological studies are constrained by limited observational and human resources, as well as by the cost of making such arduous manual inferences. As a solution to this, most astronomers have switched toward automating these tasks either through parametric regression or through citizen science projects such as  Galaxy Zoo~\citep{2008MNRAS.389.1179L}.  
Traditionally, morphological studies have been done through visual inspection. However, manual classification is limited by time, labor, and the expertise required to analyze increasingly large astronomical datasets. Solutions include community-driven classification tasks or automation. The community has increasingly adopted automated approaches, including parametric methods and citizen-science initiatives such as the Galaxy Zoo ~\citep{2008MNRAS.389.1179L}.  
Nevertheless, the emergence of large-scale astronomical surveys, including Sloan Digital Sky Survey (SDSS)~\citep {sdss}, Hyper Suprime-Cam (HSC)~\citep {HSC}, Euclid~\citep{Euclid}, and the upcoming Nancy Grace Roman Space Telescope~\citep{Spergel}, has drastically increased the volume and complexity of astronomical data, rendering purely visual classification impractical~\citep{Abraham01}.

To process this avalanche of data efficiently, machine learning and deep learning methods have become necessary tools for galaxy morphology classification \citep{Ting25}. Particularly, Convolutional Neural Networks (CNNs) have demonstrated remarkable success in automatically identifying morphological classes from imaging data. A non-exhaustive list of deep learning applications to galaxy morphology classification using data from Galaxy Zoo, SDSS, DECaLS, and other surveys can be found in \citet{Dieleman2015,Beck18,Tuccillo,Barchi,Gupta,Spindler,Bhambra21,Cao24,Luo25,Lee25,Howie25} (and references therein).

Despite these advances, most standard deep learning classifiers are trained under the assumption that test-time samples are drawn from the same ID population as the training data. In practice, astronomical survey images in general contain many objects that violate this assumption. These include rare morphologies,  merger systems, strongly disturbed galaxies, observational artifacts, or other previously unseen astrophysical phenomena. Such OOD samples are often forced into a known class with high confidence despite their disparity with training data. This behavior can lead to overlooking unusual, rare, or novel structures and reducing the reliability of downstream inference.

Recognizing these challenges, recent astronomical studies have started exploring anomaly detection and OOD-related methods. A generative approach based on variational Autoencoders (VAEs) and Wasserstein Generative Adversarial Networks (WGANs) has been used to identify mergers and tidal features in wide-field imaging surveys~\citep{2021MNRAS.508.2946S, 2020MNRAS.496.2346M}. Semi-supervised adaptation frameworks such as DeepAtroUDA have been used to align feature distributions across different surveys (e.g., SDSS, DECaLS, LSST) and across clusters of unknown classes, thereby enabling anomaly detection to verify cross-survey morphological shifts \citep{2023MLS&T...4b5013C}. OOD detection has also been used in photometric redshift estimation, where stars and quasars are treated as outliers relative to ID galaxies \citep{2022AJ....163...98L}. More recently, \citet{2024MNRAS.530.1274M} demonstrated the application of self-supervised representation learning on the Galaxy Zoo DECaLS dataset using Bootstrap Your Own Latent (BYOL) \citep{2020arXiv200607733G}, which produces low-dimensional embeddings that naturally cluster galaxies based on morphology and facilitate anomaly discovery. However, these methods focus on representation learning, anomaly identification, or domain adaptation rather than on explicit classifier-based OOD detection.

Uncertainty-aware deep learning methods have also been explored in astronomy. A Bayesian Neural Network applied to Galaxy Zoo data has produced a well-calibrated posterior label distribution while supporting active learning \citep{2020MNRAS.491.1554W, 2022MNRAS.509.3966W, 2023MNRAS.526.4768W}. However, this work focused primarily on predictive uncertainty and label efficiency rather than on evaluating OOD detection performance.


Despite the aforementioned works, there remains a need for a systematic, classifier-based OOD detector for deep neural networks that uniquely corroborates predictive uncertainty with energy/logit-aware scoring and OOD metrics on survey imagery. IsoMaxPlus (and its variants) has proven effective in the broader machine learning literature for improving entropy- and energy-based OOD detection by leveraging isotropic, distance-based logits. These logits yield higher-entropy posteriors than SoftMax, in accord with the maximum entropy principle \citep{2020arXiv200604005M, DBLP:journals/corr/abs-2105-14399}. The application of IsoMax (and its variants) combined with Bayesian approximation methods such as MC Dropout remains largely unexplored in Astrophysics. In particular, it is unclear how the behavioral aspects of IsoMaxPlus-induced logits and MC Dropout uncertainty jointly affect the energy, max-logits, and OOD-distance scores. 
Most existing works in astronomy emphasize anomaly detection \citep{LOCHNER2021100481} or domain adaptation rather than evaluating confident predictions and explicit metrics for OOD detection. Compared to prior work, our study provides one of the first systematic evaluations of IsoMaxPlus and MC Dropout for morphological OOD detection on the Galaxy Zoo DECaLS datasets, with comprehensive calibration using OOD metrics.

In this work, we propose a novel framework for OOD detection in galaxy morphology using images from the Dark Energy Camera Legacy Survey (DECaLS). We introduce a framework that combines deep learning models for feature extraction with IsoMaxPlus-based methods for identifying anomalous samples in feature spaces. In addition, we utilize MC dropout, a Bayesian approximation method, to quantify model epistemic uncertainty and further improve the detection of anomalous samples. We leverage the extensive DECaLS dataset, which contains millions of high-quality images of galaxies across multiple photometric bands,  to develop a robust and scalable OOD detection pipeline. We demonstrate the effectiveness of the proposed approach through experiments under multiple setups, supported by quantitative metrics, qualitative examples, and visualizations.

We summarize the main objectives and contributions in this work as follows:
\begin{itemize}
\item The first application of the IsoMaxPlus loss function to astronomical image classification;
\item The integration of IsoMaxPlus with Monte Carlo dropout to establish a unified framework for uncertainty quantification and OOD detection;
\item  A systematic evaluation of three configurations on the Galaxy Zoo DECaLS dataset.
\end{itemize}

This manuscript is organized as follows. Sect.~\ref{sec:dataset} describes the dataset used for the analysis. The methodology used is discussed in Sect.~\ref{sec:methodology}. The experimental design and evaluation metrics are discussed in Sect.~\ref{sec:evalmetrics}. The results and discussions can be found in Sect.~\ref{sec:results}, and the conclusions are summarized in Sect.~\ref{sec:conclusions}.



\section{Dataset}
\label{sec:dataset}
We now discuss the DECaLS dataset used for our analysis. 
DECaLS~\citep{Dey19} provides deep optical imaging in the $g$, $r$, and $z$ bands that cover approximately 9,000 deg$^2$ of the sky as part of a larger sky survey of the Dark Energy Spectroscopic Instrument (DESI) project~\citep{DESI}. DECaLS observations were made using the 4-meter Blanco Telescope at the Cerro Tololo Inter-American Observatory (CTIO), Chile, equipped with the 570-megapixel Dark Energy Camera (DECam)~\citep{Decam}. The survey achieves a $5\sigma$ point source observational depth of $g$ = 23.95, $r$ = 23.54, and $z$ = 22.50 (AB magnitudes) with the Full Wave Half-Minima (FWHM) values of  1.29, 1.18, and 1.11 arcseconds, respectively~\citep{Dey19}.

The galaxy samples used in this study are derived from the DECaLS Galaxy Zoo subset (DECaLS-GZ), which pairs DECaLS DR5 optical imaging with crowd-sourced morphological classification from the Galaxy Zoo citizen science platform \citep{2008MNRAS.389.1179L,2022MNRAS.509.3966W}. 
To enhance the samples obtained from the DECaLS archive, we cross-matched Galaxy Zoo sources with the DECaLS DR9 archive.~\footnote{\url{https://www.legacysurvey.org/dr9/description/}}
The initial targets were selected based on the spectroscopic ($0.01 \leq z \leq 0.15$) and photometric ($r_{AB} < 20.5$) cuts to prioritize the resolved morphological measurements, and later, the voter fraction from the Galaxy Zoo was taken for the final sample selection. The most appropriate criterion for selecting samples across most class types is voter percentages greater than $90\%$. However, few classes have small populations and rare morphologies, and selecting them requires a more lenient $50\%$ cutoff to meet the minimal count requirement. The reason for using such a stringent voting threshold is to pick only those samples that show significant features and are confidently voted on by fellow citizen science volunteers, as our aim is not just to classify the samples, but also to train the model in the initial phase to be confident in distinguishing   ID from OOD samples.

DECaLS imaging data can be accessed and downloaded from the public Legacy Survey data portal\footnote{\href{https://www.legacysurvey.org/}{https://www.legacysurvey.org/}}, which provides calibrated optical images from the DESI Legacy Imaging Surveys. These data form the basis for our galaxy image samples.

To preserve morphological information, the angular scale of each cutout is customized according to the apparent size and morphological class of the target object. In general, compact distant galaxies require finer angular resolution to resolve structural features, whereas a coarse scale can adequately represent more extended sources without loss of structural information. Careful selection of the angular scale is therefore essential to ensure consistent morphological constituents across a diverse galaxy population.

Table \ref{table:1} summarizes the scale factors tailored for different galaxy classes, along with the corresponding voter percentage cutoff thresholds used during sample retrieval. These thresholds ensure that only confidently classified samples are retrieved, thereby improving the reliability and representation of galaxy classes in the final dataset.

All galaxy image datasets extracted from the DESI DECaLS image viewer are initially rescaled to a uniform size of $400\times 400$ pixels, with specified scale factors. The choice of scale factors was made by manual visual inspection of randomly selected sources, accounting for variation in apparent size, intrinsic morphology, and distance (redshift). This ensures that relevant structural features were consistently captured across objects with different physical scales and distances.



\begin{table}[ht]
\begin{tabular}{ccc}
\toprule
\textbf{\textit{Class Names}}       & \textbf{\textit{Angular Resolution}} & \textbf{\textit{Volunteer vote}}  \\\midrule
& & \textbf{\textit{cutoffs}} \\

Round             & 0.1                & 0.96                  \\
Inbetween         & 0.06               & 0.95                  \\
Bars              & 0.16               & 0.75                  \\
cigar             & 0.1                & 0.95                  \\
Edge-on Bulge      & 0.1                & 0.5                   \\
Edge-on Bulge None & 0.15               & 0.86                  \\
Ring              & 0.1                & 0.6                   \\
Arms1             & 0.1                & 0.6                   \\
Arms2             & 0.1                & 0.93                  \\ 
\bottomrule
\end{tabular}
\caption{Summary of the angular resolution cuts and volunteer votes cutoff obtained from \citet{2022MNRAS.509.3966W}.}
\label{table:1}
\end{table}

\section{Methodology} 
\label{sec:methodology}
This section outlines the comprehensive experimental framework devised to rigorously evaluate both the classification efficacy and the uncertainty quantification capabilities of deep learning models applied to astronomical imaging. The motivation for this work is to evaluate model reliability under varying inputs, with a focus on ID samples that represent well-characterized astrophysical features and OOD samples that emulate previously unseen or unknown visual features. Such exploration of robustness is essential for astrophysical applications that aim to uncover unknown structures or anomalies in large-scale surveys.  

To establish a consistent baseline, all experiments utilize a common ResNet-34 model backbone, integrating separate classifier heads that follow different prediction techniques:  (1) a conventional probability-aware SoftMax classifier with cross-entropy objective function, (2) a distance-based classifier that uses class-prototype distance as logits (IsoMaxPlus), and (3) a hybrid approach (IsoMaxPlus + MC Dropout) which is based on Bayesian-inspired uncertainty estimation.

 In the final section, we describe the probabilistic approach, which replaces conventional one-hot encoding with soft target labels. This approach, rather than enforcing mutually exclusive class assignments, allows each sample to be associated with a distribution over classes, capturing intrinsic ambiguity in classes. This method promotes smoother decision boundaries in the learnable feature space and reduces overconfident predictions. In the context of galaxy morphology, where classes often overlap in their visual features, a probabilistic approach enables the model to better capture underlying morphological similarities and improve generalization across galactic classes.

\subsection{Framework overview}
We have designed a framework to separate three complementary aspects of the classification task: learning feature representations, a distance-based OOD scoring mechanism, and an uncertainty estimator. A standard ResNet-34 backbone model is used to map each galaxy image $x$ to a latent embedding $f_{\theta}(x)$. We ensure that the transformations and pre-processing steps are consistent across all techniques, allowing any performance differences in results to reflect differences among the methodologies rather than discrepancies in the feature extractor. This design is particularly appropriate for galaxy morphology, where visual similarities can blur the class boundaries and make confidence estimation more relevant.

On top of this shared representation, we compared three inference techniques: the SoftMax classifier trained with cross-entropy, which is a conventional reference point but is expected to produce overconfident posterior probabilities on unfamiliar samples, IsoMaxPlus, which replaces the standard logits with distances to learnable prototypes following the maximum isotropy principle, where the confidence is determined by geometric proximity in feature space rather than logits. We also introduce a hybrid approach to quantify the epistemic uncertainty.

\rthis{In this framework, IsoMaxPlus provides the geometric component for OOD detection by replacing the conventional logits with prototype distances, leading to an improvement in the separation between visually similar galaxy morphologies, producing a minimum distance-based novelty score in learned embedding space.}
\rthis{MC Dropout, on the other hand provides the component for quantifying epistemic-uncertainty by keeping dropout active during the inference and performing multiple stochastic forward passes. The mean value of the prediction is used for classification and the predictive variance captures the model uncertainty for ambiguous galaxies.}

\rthis{The two methods are complementary rather than redundant: IsoMaxPlus directs the class geometry and OOD separability while MC Dropout quantifies dispersion, which is especially useful for survey data containing overlapping and unlabeled morphologies.}

\begin{figure*}[h]
    \centering
    \includegraphics[width=\columnwidth, height = 0.5\textheight, keepaspectratio]{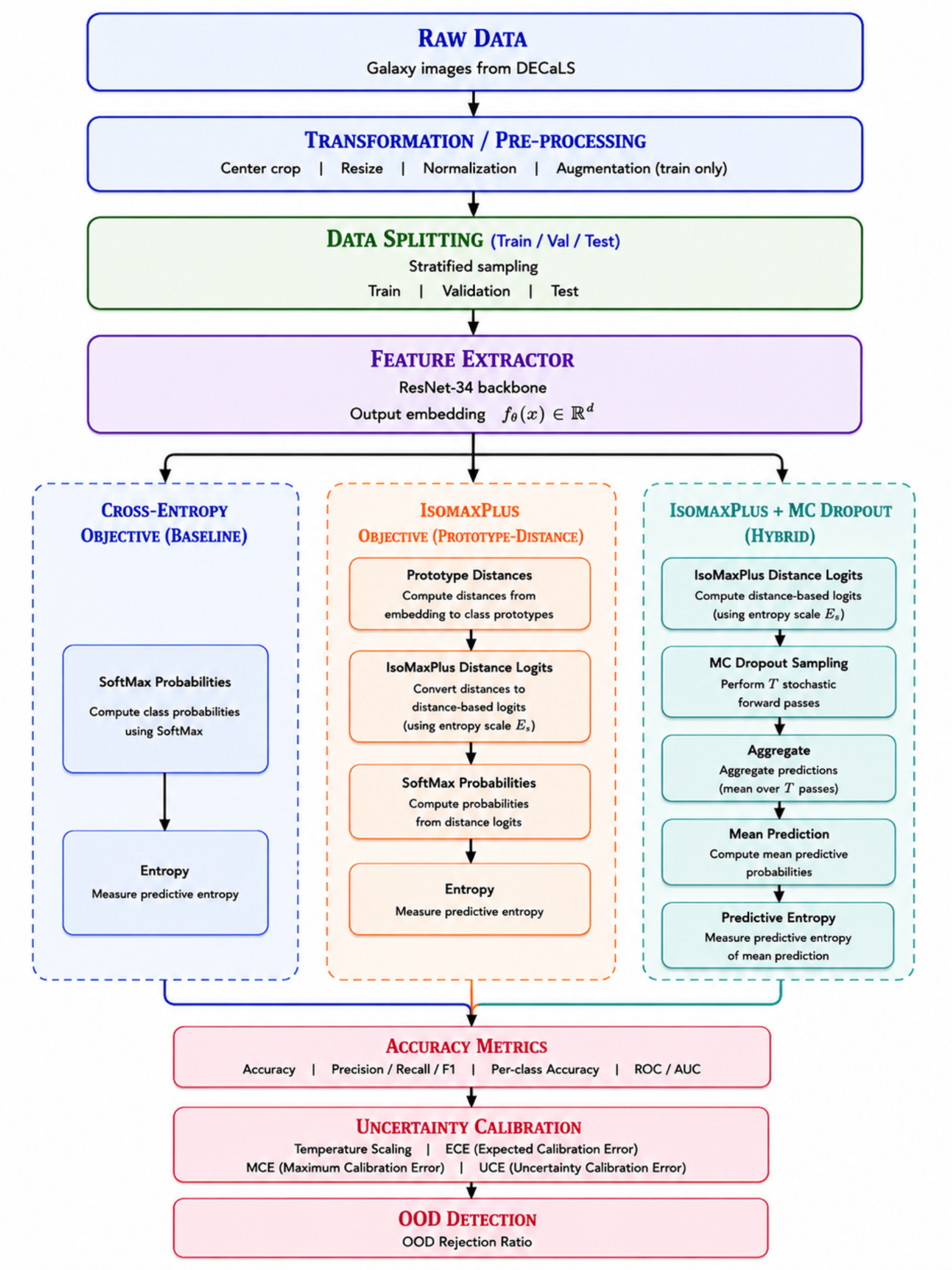}
    \caption{The above flowchart summarizes the complete experimental workflow starting from the raw samples to transformation to pre-processing, then to splitting and feature extraction steps, including the cross-entropy and IsoMaxPlus, then to the inference block, which in brief describes the steps followed, and finally to the confidence and reliability measuring metric.}
    \label{fig:schematic}
\end{figure*}

Figure~[\ref{fig:schematic}] provides a schematic summary of the workflow, from raw DECaLS images through pre-processing and feature extraction to calibration, uncertainty quantification, and OOD detection.

\subsection{Shared feature extractor : ResNet-34}
ResNet-34 is used as the shared feature extractor across all experiments. It consists of 34 layers, and is part of the family of residual networks (ResNet)~\citep{ResNet}. The fundamental design of ResNet is based on residual learning and shortcut (skip) connections to address the degradation problem that arises during training of deeper networks. For a residual block, the output can be written as:
\begin{equation}
    x_{m+1} = \sigma\!\left (x_m + F(x_m + W_m)\right)
\end{equation}
where $x_m$ and $x_{m+1}$ are the inputs and outputs of the $m-th$ unit, $\sigma(\cdot)$ denotes the activation function, $F(\cdot)$ is the residual function. For a two-layer residual block, $F = W_2 \sigma(W_1x_m)$ is the residual output, with $W_{1}$ and $W_{2}$ as convolutional weighted tensors.

\subsection{IsoMaxPlus: distance-based objective and entropy scale}

The IsoMaxPlus addresses the limitations of conventional SoftMax classifiers for OOD detection by replacing the linear logits with a distance-based objective in the learned embedding space. The Euclidean distance $d(\cdot)$ between the input and the prototype can be written as:
\begin{equation}
\label{eq:iso_distance}
    d(f_{\theta}(x), p^{i}_{\phi}) = \big\| f_{\theta}(x) - p^{i}_{\phi} \big \| 
\end{equation}
where $f_{\theta}(x)$ is the feature embedding of the input $x$, and $p^{i}_{\phi}$ is the learnable prototype (centroid) for class $i$, $||.||$ is the non-squared Euclidean distance, $d(\cdot)$ measures how close a sample lies relative to the class prototype. A sample is assigned a higher class affinity when it lies closer to the corresponding prototype and a lower affinity when it is farther away. This prototype-based geometry promotes compact class clusters and provides a natural novelty score based on the minimum prototype distance.


The IsoMaxPlus training loss is

\begin{equation}
\label{eq: isomaxplus}
    \mathcal{L_{ \text{IsoMax+} }} = -\log \left( \frac{\exp{\left(-E_{s} \big \lVert  f_{\theta}(x) - p_{\phi}^{k} \big \rVert \right) }}{\sum\limits_{j} \exp \left(-E_{s} \big \lVert  f_{\theta} (x) - p_{\phi}^{j} \big \rVert \right)} \right),
\end{equation}
where, $k$ is the true class label, $j$ is the total number of classes, $E_{s}$ denotes the entropy scale.

The Entropy scale ($E_{s}$) is a scalar quantity provided to IsoMaxPlus during the training process (Equation \ref{eq: isomaxplus}). Choosing the correct value of the entropy scale is an important step as it controls the shape of the entropy distribution; a large $E_{s}$ produces a sharper entropic distribution, while a smaller $E_{s}$ makes it flatter, i.e., the predictions are more capped. 
In practice, $E_{s}$ is chosen to control the sharpness of the training posterior and overcome overconfident predictions. Adjusting the logits-to-distance scaling during training motivates a posterior whose entropy approaches that of the maximum-entropy uniform categorical distribution, $H_{max}\simeq \log{N}$, where $N$ is the number of categories, which agrees with the principle of maximum entropy.

The inference probability does not contain the $E_{s}$ regularization, resulting in an inference probability as:
\begin{equation}
    \label{eq: prediction_probability}
    \mathcal{P_{ \text{IsoMax+} }(y = k | x)}  = \frac{\exp{\left(-\big\lVert f_{\theta}(x) - p_{\phi}^{k} \big \rVert \right)}}{\sum\limits_{j} \exp\left(-\big \lVert f_{\theta} (x) - p_{\phi}^{j} \big \rVert \right)} 
\end{equation}
The above change in Eq. \ref{eq: prediction_probability} is an important step for the Entropy maximization trick. For the classification between the ID and OOD samples, the minimum prototype distance gives a natural novelty score and can be written as follows:

\begin{equation}
\label{eq: min_distance}
    d_{\text{min}} = \min_{j} \big\lVert f_{\theta} (x) - p_{\phi}^{j}  \big \rVert
\end{equation}
Samples belonging to the ID distribution are expected to reside near one of the learned prototypes and therefore produce small $d_{\min}$ values, while unusual, ambiguous, or unseen morphologies are generally distant from all prototypes and produce larger $d_{\min}$.

\subsection{MC Dropout: Epistemic uncertainty measurement}
MC Dropout is used to estimate epistemic uncertainty, which reflects uncertainty in the model parameters; its estimation is more important when the input lies outside the training distribution. MC Dropout is a computationally efficient and widely adopted technique for estimating predictive uncertainty in deep neural networks, originally introduced by \citet{Yarin}. While dropout was initially proposed as a regularizer to mitigate overfitting during training \citep{Srivastava}, MC Dropout reintroduces this mechanism during inference to approximate Bayesian uncertainty. 

Specifically, during inference, dropout layers remain active, and the same input $x$ is propagated through the network multiple times, each time with different dropout masks. Each forward pass activates a different set of dropout masks $m^{(t)}$, yielding a stochastic ensemble of predictions that approximate an implicit variational posterior over model weights. The predictive mean summarizes the class probabilities, while the predictive variance estimates the epistemic uncertainty. 
\begin{align*}
    \hat{y}^{(t)}= &f(x; w \odot m^{(t)}), t = 1, \cdots, T ,
\end{align*}
where $w$ denotes the network weights and $\odot$ represents element-wise multiplication. This procedure can be interpreted as sampling for an approximate posterior distribution over model parameters.

The resulting predictive distribution can be summarized by its mean and variance:
\begin{eqnarray}
    \overline{y} &=&  \frac{1}{T} \sum_{t=1} ^{T}(\hat{y}^{(t)}) \\
    Var(y) &=& \frac{1}{T} \sum_{t=1} ^{T}(\hat{y}^{(t)} - \overline{y})^2
\end{eqnarray}

A key advantage of MC Dropout lies in its simplicity and scalability, as it requires no modification to model architecture and incurs only a linear increase in inference cost with the number of stochastic passes. For this reason, MC Dropout became the popular baseline for uncertainty-aware models~\citep{Yarin, Kendall}.

\subsection{IsoMaxPlus + MC Dropout}
In the previous section, we saw the importance and credibility of both IsoMaxPlus and MC Dropout. The combination of IsoMaxPlus and MC Dropout is complementary rather than redundant. IsoMaxPlus improves the geometry of the latent space with distance-based novelty score whereas MC Dropout estimates uncertainty through predictive variance. IsoMaxPlus tends to improve OOD separability in embedding space, while MC Dropout increases the predictive dispersion for uncertain inputs.
This approach offers  several advantages for a given task, one of which is hybrid scoring, which balances both  epistemic and aleatoric uncertainty. This balance makes predictions more stable and reliable when categorizing ID and OOD samples. In general, combining both techniques provides a powerful way to assess how well the model understands the data and how well the data align with the training distribution, leading to improved, more reliable and better calibrated predictions.

\subsection{Probabilistic Approach}
Conventional machine learning training processes typically utilize one-hot target labels $y \in \{0,1\}^{N}$. These labels are effective for tasks with mutually exclusive classes; one-hot labels assume absolute inter-class independence and can mask visual similarity between classes. Galaxy morphology is one such example where various classes of morphology can show resemblance with one another, such as real galaxies often exhibit mixed features (such as spirals with bars and unbarred spirals), and the volunteer classification itself contains probabilistic labels  (vote fractions)~ \citep{2008MNRAS.389.1179L, 2022MNRAS.509.3966W} instead of deterministic labels. 

Utilizing the raw vote fractions as soft labels (probabilistic labels) $q(x) = (q_1, q_2, \dots, q_N)$ where $\sum_{i = 1}^{N} q_i=1$ can reflect human uncertainty and inter-class resemblance, improving generalization and producing smoother decision boundaries for classes. This approach also helps the model learn from a more nuanced view of the data and encourages smoother decision boundaries, reducing overconfident predictions \citep{szegedy2016}. Incorporating data from multiple annotation sources, the model exhibits greater robustness to noise and is more effective at capturing the complexities of the data~\citep{raykar2010learning}.


The neural network $f_\theta$ outputs a posterior $p(x) = (p_{1}(x) \dots, p_{N}(x))$. Training with soft targets, the cross-entropy loss can be defined as:\\

\begin{equation}
    \mathcal{L}_{CE}(x) = - \sum_{k = 1}^{N} q_{k}\log{p_{k}(x)}
\end{equation}
which is negative log-likelihood under the empirical annotator distribution $q$. 
IsoMaxPlus can be aligned with soft targets and can be calculated as the IsoMax-style cross-entropy between the soft labels and the distance-based posterior as:
\begin{equation}
\begin{multlined}
    \mathcal{L}_{IsoMax} (x) = - \sum_{k = 1}^{N} q_{k}\log{p_{k}(x)}\\
    = - \sum_{k = 1}^{N} q_{k}\log \left ({\frac{\exp \left(-E_s \big\lVert f_{\theta}(x) - p_{\phi}^{k} \big \rVert \right)}{\sum_{j}\exp \left(-E_s \big \lVert f_{\theta}(x) - p_{\phi}^{j}  \big \rVert \right)}} \right)
\end{multlined}
\end{equation}
The above equation reduces to the standard equation \ref{eq: isomaxplus} when the soft targets are replaced with the usual one-hot-encoded vector. Training with empirical vote-fraction targets (e.g., Galaxy Zoo DECaLS) enables the model to understand real-data annotator uncertainty and avoid adopting an overly confident decision boundary.

\section{Experimental design and Evaluation metrics}
\label{sec:evalmetrics}
This section describes the experimental framework implemented to assess classification performance, calibration quality, and uncertainty estimation for the proposed techniques. The goal is to highlight differences in the architecture's behavior across various model setups for both ID and OOD setups, which is essential for automated analysis of astronomical data. 



To achieve this, we design a series of controlled experiments involving multiple neural network architectures built on the ResNet-34 backbone, each incorporating different techniques for improving predictive confidence, calibration, and generalization. This includes the use of softmax for the baseline estimation, IsoMaxPlus for its unique OOD detection capability, and MC Dropout for the uncertainty estimation. Additionally, we implemented a hybrid approach that combined IsoMaxPlus with MC Dropout to assess potential synergistic effects on both prediction accuracy and uncertainty quantification. The experiments are designed to assess how well different model configurations classify astronomical data and to quantify their predictive uncertainty and robustness to OOD data.

\subsection{Data splitting and Pre-processing}
Following the data acquisition and curation procedures described in Section~\ref{sec:dataset}, the dataset was partitioned into three mutually exclusive subsets: training, validation, and testing. A stratified split was used to preserve the class distribution within the training and validation sets. The final test set was constructed independently and contains samples with different class distributions, providing an unbiased estimate of generalization performance. 

\rthis{The training and validation sets are split into $75:25$ to preserve the class proportion. Further to limit the class dominance and maintain a balanced training process, the sum of training and validation set is maintained at a maximum of $3000$ samples per class.}
\rthis{The test set was constructed independently and kept fixed throughout the experiments. The test set contains the remaining samples, which are not used during training and also include the OOD samples.}
\rthis{Accordingly, the class percentage reported in Table~\ref{tab:counts} are computed with respect to the total number of samples within each respective sets.}  

A summary of the composition of the data set, including class counts in the training, validation and test sets, is provided in Table~\ref{tab:counts}. This table highlights the distributional differences between the splits, ensures transparency, and provides details for reproducibility.

Following the splitting process, pre-processing, and image-level augmentation are applied to make the training process invariant to spatial position and orientation. Initially, to handle the varying pixel magnitudes, each image was normalized by rescaling the pixel values within the range of $[0,1]$. This normalization ensures a consistent input scale across samples, stabilizes gradient optimization, and accelerates convergence.

To further enhance robustness, data augmentation was applied in the training set. Including random rotations, horizontal and vertical flips, and color jitter increases the effective diversity of the training set by introducing invariance. With these transformations,  realistic observational variations can be added to the astronomical imaging, including changes in galaxy orientation. All augmentations and normalizations have been implemented using the {\tt torchvision.transforms} module in  {\tt Pytorch}. Only normalization was applied to the test and validation sets. No augmentation processes were applied to the validation or test sets to keep the model selection and final evaluation unbiased.

\begin{table*}[h]
\centering
\caption{\rthis{Class distribution across training, validation, and testing set along with the OOD classes and counts. The training set was made selecting a maximum of 3000 samples per class. Percentage for the training and validation sets are computed with respect to the total number of samples in respective split. The testing set is exclusive samples set unseen during the overall training progress.}}
\label{tab:counts}
\resizebox{\columnwidth}{!}{
\begin{tabular}{|lccccc|}
\toprule
\textbf{Class}                                             & \textbf{\textit{Training}} & \textbf{\textit{Training(\%)}} & \textbf{\textit{Validation}} & \textbf{\textit{Val (\%)}} & \textbf{\textit{Testing}} \\ 
\midrule
Round                                                      & 2382           & 12.96                & 618     & 13.45          & 13236         \\
Inbetween                                                  & 2427           & 13.20                & 573     & 12.47          & 10561         \\
Arms1                                                      & 645            & 3.51                 & 155     & 3.37           & 368           \\
Arms2                                                      & 2367           & 12.88                & 633     & 13.77          & 12075         \\
Bars                                                       & 1096           & 5.96                 & 234     & 5.09           & 767          \\
Ring                                                       & 2402           & 13.07                & 598     & 13.01          & 1282          \\
Cigar                                                      & 2412           & 13.12                & 588     & 12.79          & 10405         \\
Edge-on Bulge None                                         & 2276           & 12.38                & 572     & 12.45          & 1299          \\
Edge-on Bulge                                              & 2375           & 12.92                & 625     & 13.60          & 2027          \\
\midrule
\textbf{Total}                                             & \textbf{18382} & \textbf{100.00} & \textbf{4596} & \textbf{100.00} & -- \\
\midrule
\multicolumn{6}{|c|}{\textbf{\textit{Additional Test Classes (OOD)}}}                             \\
\midrule
Merger (OOD)                                               & --             & --                   & --      & --             & 5756          \\
Irr (OOD)                                                  & --             & --                   & --      & --             & 4098          \\
Bulge Round(OOD)                                           & --             & --                   & --      & --             & 3828          \\
Merger Dist (OOD)                                          & --             & --                   & --      & --             & 1324         \\
\bottomrule
\end{tabular}
}
\end{table*}

\subsection{Entropy score and baseline objective}
In addition to classification performance, we evaluate the model's predictive uncertainty using entropy. In information theory, the degree of uncertainty associated with a probability distribution can be quantified by entropy. Similarly, in deep neural networks, entropy serves as a proxy for understanding the model's confidence in predictions and predictive uncertainty. Given a model's predictive probability distribution $P(y_{i}|x)$ for input $x$ across the $N$ classes, predictive entropy can be defined as: 
\begin{equation}
\begin{aligned}
\label{eq:es}
H(x) = -\sum_{i=1}^{N} p(y_{i}|x) \log{p(y_{i}|x)}
\end{aligned}
\end{equation}

Lower entropy indicates more confident predictions and a peaked entropy distribution, whereas higher entropies tend to have flatter distributions and a more uncertain predictive distribution. This property makes entropy a useful baseline for OOD detection, since the OOD samples often yield more uncertain predictions than ID samples.


In our experiments, we selected cross-entropy as the baseline evaluation metric to provide a consistent, interpretable reference point for model performance. Cross-entropy is widely used for classification tasks because it directly measures the discrepancy between predicted probabilities and true labels. In addition, it not only quantifies classification error but also has a solid theoretical foundation as the negative log-likelihood under a categorical distribution, as discussed by \citet{Goodfellow-et-al-2016}.

\subsection{Technical details of implementation}
We trained our model using stochastic gradient descent (SGD) with a batch size of $256$, selected to fully utilize available GPU memory and stabilize gradient estimates at larger batch sizes. A momentum factor of $0.5$ was chosen to accelerate convergence and reduce oscillations during optimization. The initial learning rate is set to $1e-3$, with a one-cycle learning rate scheduler provided by {\tt PyTorch}, for a total of $140$ epochs. This learning rate was chosen to achieve rapid initial convergence while gradually refining in later epochs, thereby improving both convergence and generalization. To prevent overfitting and ensure optimal model selection, an early stopping criterion with a patience of $30$ epochs is used. 

A weight-decay (L2 regularization) factor of $1e-4$ was applied to penalize large weights and improve generalization. The weights are initialized as proposed by \citet{2016cvpr.confE...1H}. The entropy scale factor for IsoMaxPlus loss was set to 10 to balance class separation and confidence calibration. We also adopted batch normalization after the convolutional layer and, before the activation function, as recommended by \citet{2016arXiv160305027H}.

The entire framework has been developed using Python with Pandas, Scikit-learn for data handling and pre-processing, NumPy, and PyTorch for model development and training. On average, the time to train a single model was $\approx 3$ hours on a NVIDIA GeForce GPU with 24 GB of memory running under the CUDA 11.4 environment.

\begin{table}[]
\centering
\resizebox{\columnwidth}{!}{
\begin{tabular}{|ccc|}
\toprule
\label{tab:parameters}
\textbf{Parameter}   & \textbf{Value}                 & \textbf{Notes}        \\
\bottomrule
\bottomrule
\multicolumn{3}{|c|}{\textbf{\textit{Data \& Splitting}}}                      \\ 
\bottomrule
Train / Val Split    & 75\% \& 25\%                   & Stratified sampling   \\
Train Augmentation   & Rot, Flip, Crop                & Train only            \\
Validation Aug       & None                           & Only normalization    \\ 
\toprule
\multicolumn{3}{|c|}{\textbf{\textit{Training Setup}}}                                  \\
\bottomrule
Device               & CUDA                           & GPU training          \\
Batch Size           & 256                      & Sample counts         \\
Workers              & 11                             & Data loading          \\
Epochs               & 140                      & Training span         \\
Optimizer            & SGD                            & Standard optimization \\
Momentum             & 0.5                     &                       \\
Weight Decay         & 10\textasciicircum{}\{-4\}     & L2 regularization     \\
Gradient Clip        & 0.8                            & Stability             \\
\toprule
\multicolumn{3}{|c|}{\textbf{\textit{Learning Rate}}}                                    \\
\bottomrule
LR                   & 10\textasciicircum{}\{-3\}     & Learning Progress     \\
Scheduler            & OneCycleLR                     & Cosine annealing      \\
Cycles               & 1                              & Custom cycles         \\ 
\toprule
\multicolumn{3}{|c|}{\textbf{\textit{Regularization \& Control}}}                        \\ 
\bottomrule
Early Stopping       & 30                       & Patience              \\
Min Delta            & 5 * 10\textasciicircum{}\{-5\} & Convergence threshold \\ 
\toprule
\multicolumn{3}{|c|}{\textbf{\textit{Uncertainty \& OOD}}}                               \\ 
\bottomrule
Entropy Scale (E\_s) & 10  (optimized recommendation)                          & IsoMaxPlus            \\
\bottomrule
\end{tabular}
}
\caption{Summary of all the related parameters involved for the experimentation, including the data splitting details, augmentation-related information, and the training parameters, along with MC Dropout inference parameters}
\end{table}

\subsection{Interpretability with Grad-CAM}
To interpret the model’s predictions, we additionally used Gradient-weighted Class Activation Mapping (Grad-CAM). Grad-CAM is a powerful visualization technique that identifies and highlights the specific regions of an image that most strongly influence a particular classification decision. By analyzing the gradients flowing into the last convolutional layer of a neural network, Grad-CAM produces a heatmap indicating which areas of the image were most significant in driving the model's prediction. In the context of galaxy morphology, this allows us to evaluate whether the model attends to physically meaningful structures, such as spiral arms, bulges, bars, tidal features, or other features, rather than to spurious background or artifacts. This allows for a deeper understanding of how the model interprets visual data, shedding light on the features it considers important in making its decision. We use these visualizations as a complementary diagnostic to the quantitative OOD metrics, particularly for inspecting representative ID and OOD examples.

\subsection{Evaluation metrics}
\label{sec:evalMetric}
We use Accuracy and Macro-F1 score as primary indicators for classification performance~\citep{Bethapudi}.
Accuracy measures the fraction of correctly classified samples and is a global indicator of classification performance:
\begin{equation}
    \mathrm{Accuracy} = \frac{1}{N} \sum_{i=1}^{N} \mathbf{1}(\hat{y}_i = y_i),
\end{equation}

Accuracy is easy to interpret but can be misleading under class imbalance. For this reason we also use {\tt Macro-F1}, defined as the unweighted average of per class \textit{F1} score computed separated for each class:
\begin{align}
    \mathrm{F1}        &= \frac{2 \cdot \mathrm{Precision} \cdot \mathrm{Recall}}{\mathrm{Precision} + \mathrm{Recall}}.
\end{align}
{\tt Macro-F1} treats all the classes equally, making it an essential metric in multi-class classification where class imbalance is persistent.

\subsection{Calibration and Uncertainty metrics}
For tasks involving uncertainty quantification and OOD detection, relying solely on classification metrics may introduce bias and not accurately reflect the model's confidence in its predictions. A model may achieve high accuracy even when its confidence scores are still poorly calibrated, which can lead to ambiguity in unseen data. Therefore, we also evaluate an uncertainty-aware metric that explicitly measures the alignment between predictive and empirical correctness. Uncertainty calibration metrics are well-suited for tasks involving real-world examples, such as astronomical imaging.
A well-calibrated model produces empirically aligned predictions. We adopted both confidence-based and entropy-based calibration measures to provide a comprehensive evaluation of uncertainty quality.
\begin{itemize}
\item \textit{\textbf{Expected calibration error}}\\
The Expected calibration error (ECE) measures the average discrepancy between the predicted confidence and observed accuracy across a set of confidence bins. For $M$ bins: 
\begin{equation}
    ECE = \sum_{m=1}^{M} \frac{n_{m}}{N} \left|a_{m} - \hat{p}_{m}\right| ,
\end{equation}
where,\\
$n_m = \left| B_m\right|$ is the number of samples in bin $m$\\
$a_m$ is the empirical accuracy of bin $m$\\
$\hat{p}_{m}$ is the average confidence of  the prediction in bin $m$\\

ECE provides a global scalar measure of calibration quality but hides details of the localized calibrated error due to its averaging nature.
\item \textit{\textbf{Maximum calibration error}}\\
MCE represents the maximum discrepancy between the accuracy and prediction confidence across all bins:
\begin{equation}
    MCE = \max_{m \in \{1, \ldots, M\}} \left|a_{m} - \hat{p}_{m}\right|
\end{equation}

MCE complements ECE by highlighting models that are extremely mis-calibrated across all bins.

\item \textit{\textbf{Uncertainty calibration error}}\\
To address the limitations of confidence-based calibration metrics in multi-class setups, we also compute the Uncertainty calibration error (UCE), which replaces a prediction's confidence with its entropy as the uncertainty measure.
\begin{equation}
\mathrm{UCE} = \sum_{m=1}^{M} \frac{|B_m|}{N} \left| \mathrm{\epsilon}(B_m) - \mathrm{\varepsilon}(B_m) \right|,
\end{equation}
where, $\mathrm{\epsilon(B_m)}$ is the empirical error rate in bin $m$ and, $\mathrm{\varepsilon(B_m)}$ is the average normalized entropy of prediction in bin $m$ for $B_{m}$ the entropy in m-th bin. UCE provides a more refined assessment of uncertainty calibration by directly comparing the prediction uncertainty with the observed error frequency, making it suitable for OOD-aware detection. Further details on UCE can be found in \cite{laves2021}.

\end{itemize}

\section{Results and Discussion}
\label{sec:results}
In this section, we compare the classification, calibration, and OOD detection capabilities of the three proposed variants evaluated on the DECaLS galaxy morphology dataset. A standard cross-entropy model is used as the baseline, whereas IsoMaxPlus and hybrid \textit{IsoMaxPlus+MC Dropout} models utilize uncertainty-aware capabilities. We report standard classification metrics as discussed in Section~\ref{sec:evalMetric}, including accuracy and macro-F1 for classification, entropy and minimum distance for OOD, whereas the calibration is measured using \textit{ECE}, and \textit{MCE}. Due to class imbalance in the dataset, macro-F1 scores reflect the minority class information equally compared to that of the majority.


\subsection{Results}
The study evaluates nine morphological classes: Round, In-between, Bars, Cigars, Edge-on bulge, Edge-on without bulge, Arms-1, Arms-2, and Rings. The general classification results are summarized in Table~\ref{tab:avg_accuracy_f1}. For the cross-entropy baseline, the model achieves an accuracy of $0.9748 \pm 0.0011$ and a macro-F1 score of $0.9435 \pm 0.0033$. IsoMaxPlus improves these values to $0.9838 \pm 0.0007$ and macro-F1 to $0.9612 \pm 0.0019$, respectively. This improvement suggests that the distance-based objective produces a more balanced decision boundary in feature space, translating to better class separability and more balanced performance across classes. This IsoMaxPlus objective also affects the average performance of the less frequent and morphologically ambiguous classes.

Following \cite{2022A&C....3900557L}, who implemented Analysis of Variance (ANOVA) to determine whether the differences between machine-learning models correspond to statistically significant changes in the target estimates, we conducted a comparison of the three configurations across six runs, each trained with different initial conditions. This was achieved using repeated-measures ANOVA tests, with the same test set applied for all assessments. The results indicated no significant statistical dependence on the method for accuracy, as evidenced by $F(2,10) = 3.01$ and $p$-value $= 0.095$. Similarly, for the macro-F1 score, ANOVA revealed no significant dependence, with $F(2,10) = 2.41$ and $p$-value $= 0.1395$.


The per-class F1 scores in Table~\ref{tab:f1_comparison} show that the improvement is not uniform across all classes. The most noticeable gains are therefore observed in the minority classes, while the dominant classes remain saturated across all three settings. This pattern is consistent with the Galaxy Zoo DECaLS labels, which encode morphological uncertainty via vote fractions rather than imposing perfectly sharp class boundaries. Furthermore, the residual errors are not random; they are concentrated in classes that are difficult to distinguish from one another because of visual similarities, projection effects, or weak structural contrast. 


This is also reflected in the per-class recall and precision values, which follow the same trend as the F1-score. In particular, the IsoMaxPlus model improves performance on minority and structurally ambiguous classes more effectively than the cross-entropy baseline, indicating that the learned representation is better aligned with the dataset's morphological structure. 

Figures~(\ref{fig:ce_mc}, \ref{fig:iso_mc}, \ref{fig:iso_mc_mc}) show the distribution of predictive entropy $H(x)$ (Eq. \ref{eq:es}) and minimum distance $d_{min}$ to the prototype centroid (Eq. \ref{eq: min_distance}) for ID and OOD samples under the three training setups. Predictive entropy is a standard uncertainty summary that measures how diffuse the posterior prediction is; the minimum distance captures the geometric proximity of a sample to the nearest learned class prototype. The combination thus provides complementary views of confidence: entropy is probabilistic while minimum distance is geometric. 

In the cross-entropy baseline setup, even though the OOD samples exhibit slightly higher entropy on average, the distribution reflects significant overlap with the ID samples, indicating weak uncertainty in calibration. Furthermore, the distance distribution shows an inconsistent geometric structure, in which OOD samples are mapped to nearby regions relative to the class prototypes. This arises from the anisotropic behavior of the SoftMax objective function, which is known to produce overconfident predictions even for OOD inputs.

In contrast, the IsoMaxPlus produces a substantially more structured separation in both entropy and distance space. ID samples are concentrated in low-entropy, low-distance regions, while OOD samples shift towards higher entropy and larger distances from all prototypes. The OOD entropy distribution moves from a mean of approximately $-0.358$ in the baseline to about $-0.22$ under IsoMaxPlus, indicating that the model is less confident in OOD inputs, and is therefore more consistent with the uncertainty expected for anomalous objects. The hybrid model preserves this separation while making the uncertainty estimates more stable across stochastic forward passes.

Complementing the aforementioned metrics, we also inspect the learned embedding space using UMAP projection of the latent features. This visualization gives a direct view into how the ResNet-34 model organizes the galaxy morphology under different proposed inference strategies. The UMAP shows that the baseline has a relatively broader overlap between neighboring regions. 

Comparing the entropy distributions, IsoMaxPlus shows a more apparent distinction between the ID and OOD entropy peaks, which significantly reduced overlap relative to the cross-entropy counterparts. This reduced overlap suggests that the IsoMaxPlus + MC Dropout samples provide a more reliable and confident approach to identifying OOD, significantly boosting the model's detection capabilities. 

The calibration analysis further demonstrates the benefits of uncertainty-aware inference. Using 20 equally spaced confidence bins, the ECE decreases from $0.0095 \pm 0.001$ for baseline to $0.0033 \pm 0.0005$ for hybrid. This diminishing value indicates a substantial improvement in predictive confidence relative to empirical accuracy. While IsoMaxPlus improves the geometric separation between ID and OOD samples, the addition of MC Dropout further reduces overconfidence by incorporating epistemic uncertainty through stochastic inferences.

Figure~\ref{fig:gradcam_correct} shows representative Grad-CAM maps for galaxies that have been accurately classified. In these maps, the network emphasizes morphologically significant regions such as the bulge, spiral arms, bar structure, tidal features, and the galaxy's prominent central core, depending on the class. This indicates that the classifier is relying on relevant physical features rather than irrelevant background to make its predictions. The correlation between the activation areas and the visible morphology provides qualitative evidence supporting the reliability of the learned representations.

Figure~\ref{fig:gradcam_wrong} shows the Grad-CAM maps for misclassified galaxies. In these maps, the activation patterns tend to be more diffuse, fragmented, or misaligned with the most informative morphological structures. In many instances, the network appears to focus on ambiguous regions, low-surface-brightness areas, surrounding companions, or imaging artifacts rather than on the dominant morphological signature of the galaxy. This behavior suggests that the misclassifications may arise from intrinsically subtle morphology, class overlap, low signal-to-noise ratio, or a distribution shift relative to the training set. 

\begin{table*}[ht]
\centering
\begin{tabular}{|l c c c|}
\toprule
Metric & CE & Iso & Iso+MC \\
\midrule
Accuracy & 0.9748 $\pm$ 0.0011 & 0.9838 $\pm$ 0.0007 & 0.9838 $\pm$ 0.0010 \\
F1 Score & 0.9435 $\pm$ 0.0033 & 0.9612 $\pm$ 0.0019 & 0.9590 $\pm$ 0.0022 \\
\bottomrule
\end{tabular}
\caption{Comparison between various types of runs with average of accuracy and macro-F1 scores along with $1\sigma$ uncertainties, the average is obtained over six different runs with different initialization seeds and over 33 \% of data being considered for a single run}
\label{tab:avg_accuracy_f1}
\end{table*}

\begin{table}[ht]
\centering
\begin{tabular}{|c c c c|}
\toprule
Class & F1 (CE) & F1 (Iso) & F1 (Iso +   MC) \\
\midrule
0 & 0.998 $\pm$ 0.000 & 0.998 $\pm$ 0.000 & 0.998 $\pm$ 0.000 \\
1 & 0.999 $\pm$ 0.000 & 1.000 $\pm$ 0.000 & 1.000 $\pm$ 0.000 \\
2 & 0.784 $\pm$ 0.019 & 0.820 $\pm$ 0.011 & 0.793 $\pm$ 0.012 \\
3 & 0.958 $\pm$ 0.002 & 0.971 $\pm$ 0.001 & 0.971 $\pm$ 0.002 \\
4 & 0.928 $\pm$ 0.008 & 0.963 $\pm$ 0.006 & 0.967 $\pm$ 0.005 \\
5 & 0.881 $\pm$ 0.006 & 0.931 $\pm$ 0.004 & 0.932 $\pm$ 0.007 \\
6 & 0.991 $\pm$ 0.001 & 0.994 $\pm$ 0.001 & 0.995 $\pm$ 0.001 \\
7 & 0.968 $\pm$ 0.005 & 0.981 $\pm$ 0.003 & 0.984 $\pm$ 0.001 \\
8 & 0.986 $\pm$ 0.001 & 0.992 $\pm$ 0.002 & 0.992 $\pm$ 0.001 \\
\bottomrule
\end{tabular}
\caption{Comparison of Per-Class F1 Scores (along with $1\sigma$ uncertainties) across three different experimental setups with similar conditions mentioned in Table \ref{tab:avg_accuracy_f1}}
\label{tab:f1_comparison}
\end{table}

\begin{table}[ht]
\centering
\caption{Class-wise threshold values for various techniques for detecting TNR@95TPR values}
\begin{tabular}{|c c c c|}
\toprule
Class & CE & IsoMaxPlus & IsoMaxPlus + MC \\
\midrule
0 & 0.998 & 0.990 & 0.989 \\
1 & 1.000 & 0.992 & 0.991 \\
2 & 0.556 & 0.556 & 0.541 \\
3 & 0.765 & 0.901 & 0.900 \\
4 & 0.749 & 0.762 & 0.853 \\
5 & 0.727 & 0.705 & 0.690 \\
6 & 0.986 & 0.981 & 0.981 \\
7 & 0.834 & 0.884 & 0.913 \\
8 & 0.989 & 0.972 & 0.978 \\
\bottomrule
\end{tabular}
\label{tab:tnr_tpr95_methods}
\end{table}

In Table~\ref {tab:tnr_tpr95_methods}, the class-wise threshold values are provided to identify the probable In and OOD samples. These values clearly show that choosing a threshold to detect the overall True Negative Rate (TNR) at a $95\%$ True Positive Rate (TPR) does not yield the most significant detections. The improvement in the values of minority classes for the IsoMaxPlus and IsoMaxPlus + MC techniques again demonstrates their significance in detecting OOD samples at higher thresholds.

\subsection{Discussion}
We examined the effect of modified ResNet models trained with the IsoMaxPlus loss function, as well as a hybrid IsoMaxPlus+MC Dropout approach, for OOD galaxy classification on the Galaxy Zoo DECaLS dataset, using cross-entropy as the baseline. The results show that the proposed method exhibits better calibrated and more robust predictive behavior compared to the baseline, particularly for OOD detection.

IsoMaxPlus achieved notable improvements in macro-F1 and accuracy particularly in classification performance on minority classes. No explicit class-imbalance corrections were applied, as we aim to identify the method that is more stable and effective at detecting samples under identical training conditions. This implies that the summarized differences can be attributed to the loss function and uncertainty rather than to resampling and reweighting.  

\begin{figure*}[h]
    \centering
    \subfloat[\textbf{Cross-Entropy baseline}\label{fig:ce_mc}]{
    \includegraphics[width=0.9\linewidth]{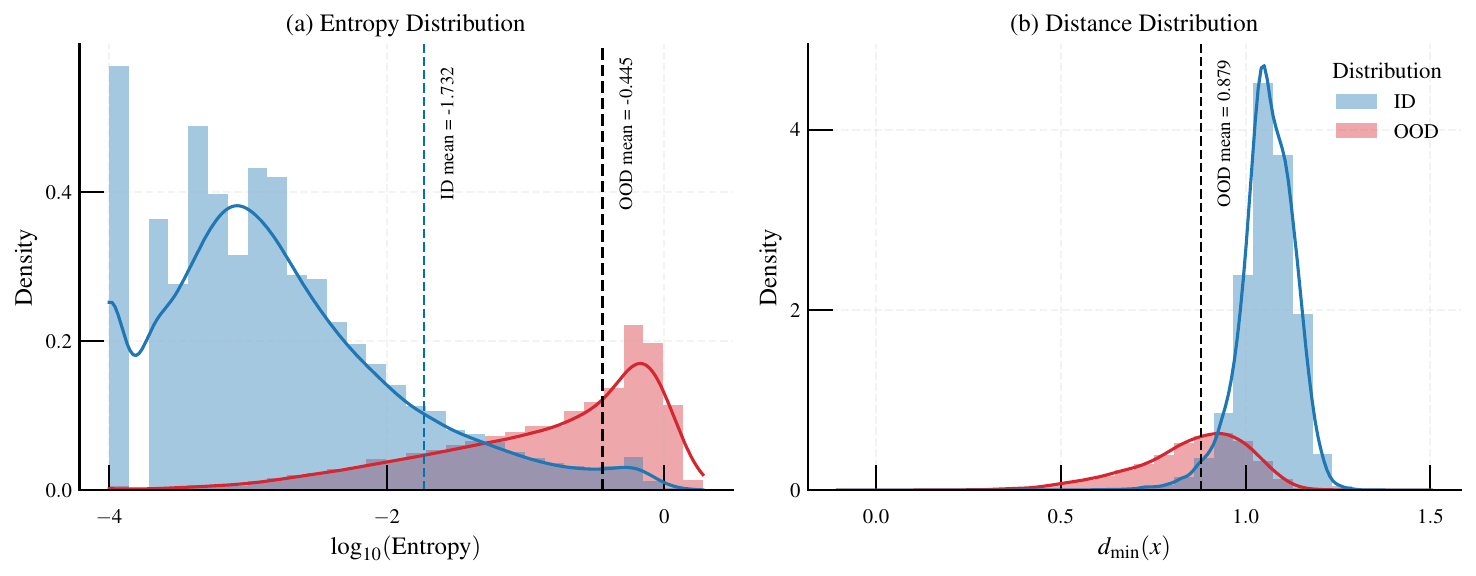}}

    \subfloat[\textbf{IsoMaxPlus}\label{fig:iso_mc}]{
    \includegraphics[width=0.9\linewidth]{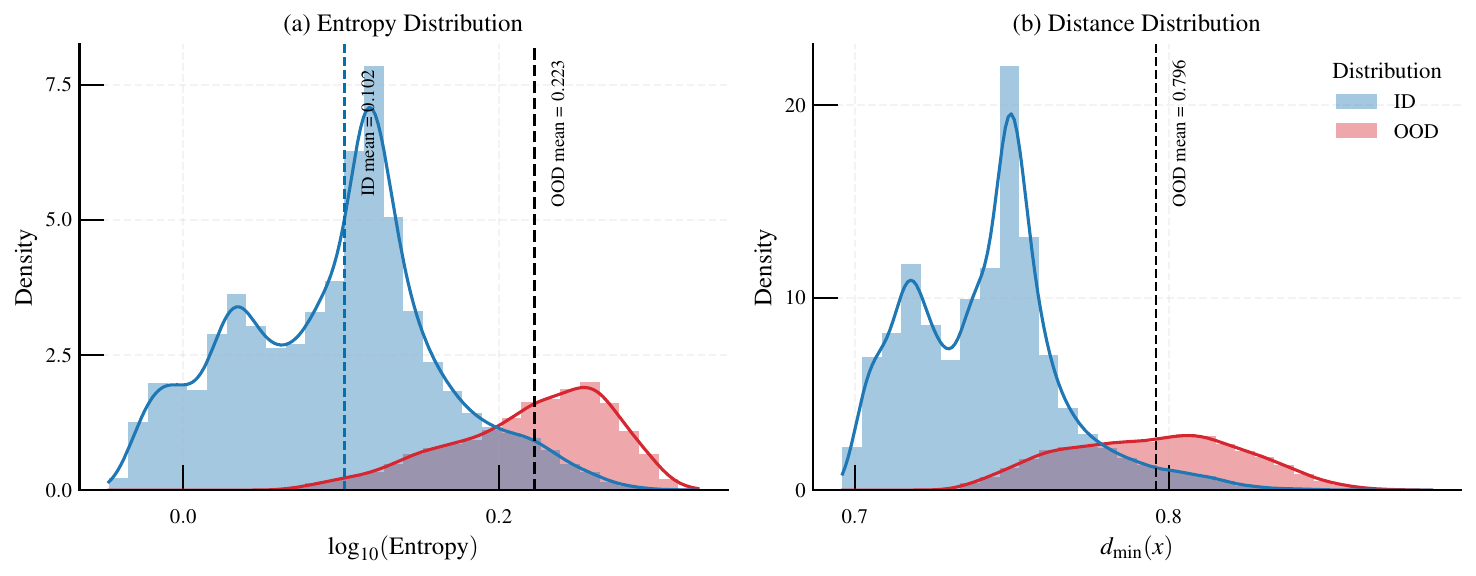}}
    
    \subfloat[\textbf{Hybrid}\label{fig:iso_mc_mc}]{\includegraphics[width=0.9\linewidth]{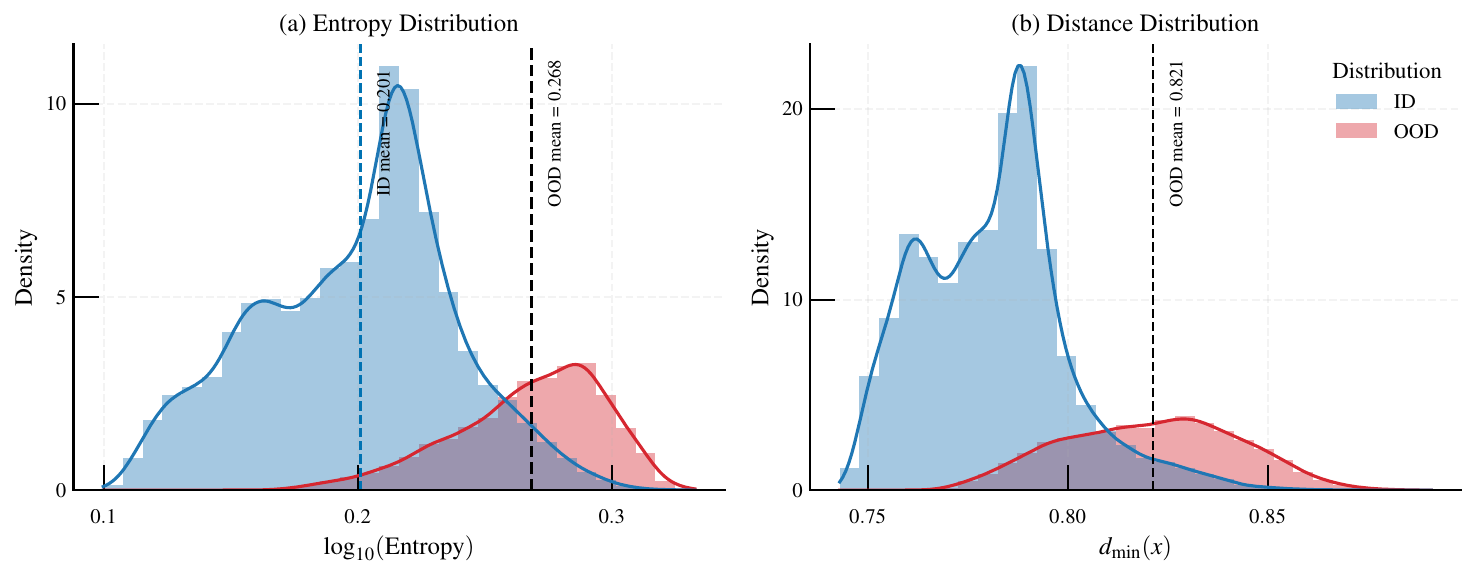}}
    \caption{Comparison of entropy distribution of ID and OOD scores for (i)the cross-entropy baseline, (ii) IsoMaxPlus, and (iii) the hybrid. Each panel shows the logarithmic entropy distribution, and the minimum distance score $d_{min}(x)$ in (a, b). The distributions of ID and OOD samples are represented by red and blue, respectively, and the solid curve shows the corresponding smoothed density estimates. The vertical dashed lines indicate the mean score of each population. Entropic distribution of the model with IsoMaxPlus and MC Dropout in logarithmic scale with ($0.268$) mean for OOD, the predictions are supposed to be more stable as they are iterated over 50 forward passes applying MC Dropout at inference}
\end{figure*}

There is another aspect to consider: the classes under review have a noticeable resemblance in their visual morphology, which can affect the confidence and entropy scores, as well as the calculated minimum distances. Nevertheless, the concentration of peaks in the entropy plots suggests that the model separates familiar and unfamiliar inputs more effectively in IsoMaxPlus than in the baseline model. Although some overlap in the distributions is still expected, the separation is considerably more reliable.
Observing the entropy distribution plots, the IsoMaxPlus model's output entropy distribution, as shown in Figure~(\ref{fig:iso_mc}), displays a clearer bimodal structure than the baseline entropy distribution in Figure~(\ref{fig:ce_mc}).

The observed bimodal entropy structure in IsoMaxPlus (cf. Figure~\ref{fig:iso_mc}) is in accordance with the feature space isotropy. By enforcing a distance-based representation, the method promotes a more uniform distribution of features in the latent space, reduces directional biases, and improves the separability of ID and OOD inputs. 
This geometric restructuring leads to distinct peaks and can be seen in the  bimodal shape for ID and OOD.
When viewed through the lens of maximum entropy principle, the bimodal  peaks are the result of confident predictions for ID samples, which peak in the lower entropy region of the plot. In contrast,  the less confident or OOD samples occupy a flatter distribution  at higher entropy,  reflecting the greater uncertainty associated with these predictions.

Similarly, Figure \ref{fig:iso_mc_mc} shows that the MC Dropout variant preserves the overall distribution with more stable sampling across 50 forward passes, and this technique is widely used because it introduces uncertainty in inference without changing the base architecture. From a Bayesian perspective, MC Dropout approximates posterior predictive uncertainty by sampling different sub-networks at inference, resulting in a reduction in the calibration error. \rthis{The hybrid approach produces more nuanced estimates for OOD inputs because it relies on the predictive distribution, which is sensitive to epistemic uncertainty, rather than on a single deterministic SoftMax output.}

In Figure \ref{fig:ce_mc}, the overlap is much broader, with a much sparser distribution of the entropy values, indicating weak calibration and a weak capability to detect OOD. This can also be viewed as a justification for the overconfident softmax outputs of the baseline technique.

Examining the mean of the entropic distribution in various figures further supports the selection of the best models for OOD detection. Other metrics, such as class-wise TNR@TPR95 values, further confirm that the model distinguishes known samples from novel samples more reliably. Also, when looking at the uncertainty calibration perspective, we see a similar nature of outcomes in determining the advantages of IsoMaxPlus and MC Dropout techniques in the form of lower ECE values, again suggesting the ability of these techniques in generating reliable, stable, and more robust predictions, which is beneficial for large-scale astronomical survey classification tasks.

\section{Conclusions}
\label{sec:conclusions}
We have developed an uncertainty-aware deep learning technique for automating galaxy morphology classification, combining distance-aware learning with Bayesian uncertainty estimation. Using a shared ResNet-34 backbone feature extractor, we evaluated three inference techniques: a conventional SoftMax cross-entropy baseline, IsoMaxPlus prototype-based classification, and a hybrid framework (IsoMaxPlus + MC dropout) for predictive uncertainty estimation and OOD detection.

The proposed IsoMaxPlus function improves the geometric feature space by replacing the conventional logits with learnable class prototypes. This improves galaxy classification by reducing the intra-class spread and increasing the separation between ID and OOD samples in both the entropy and minimum-distance metric. Compared to the baseline, IsoMaxPlus achieves improved classification performance, with accuracy reaching $0.9838 \pm 0.0007$ and macro-F1 reaching $0.9612 \pm 0.0019$, along with the added capability of detecting  OOD samples. The addition of MC Dropout further improves calibration by approximating Bayesian uncertainty through repeated stochastic forward passes during inference, preserving accuracy while reducing the ECE from $0.0095 \pm 0.001$ to $0.0033 \pm 0.0005$.

Furthermore, the shift in the OOD entropy distribution and the increase in the minimum distance demonstrate that the proposed framework can identify samples that are weakly represented by the training distribution.


These enhancements are consistent with the method's design: IsoMaxPlus uses distance-based logits and an entropy-scale regularization to encourage more structured feature-space separation, whereas MC Dropout approximates Bayesian predictive uncertainty by performing multiple stochastic forward passes during inference.

We have observed that misclassified galaxies exhibit significantly higher predictive entropy and lower minimum-distance scores in the IsoMaxPlus latent space, enabling astronomers to quantify confidence and identify unreliable predictions. This ability is essential in domains such as astronomical morphological classification, where vast amounts of data are generated in large-scale surveys, making manual verification impractical. Additionally, the minimum distance score provides an effective method for bifurcating ID samples from OOD data, including observational artifacts, novel morphological shapes, and merger remnants, without requiring explicit training on such outlier examples.

The proposed framework is particularly relevant for current and upcoming astronomical surveys, including Vera Rubin LSST, EUCLID, Roman, and DESI-DECaLS itself, where data volume is rapidly growing. In such cases, automation is gradually replacing manual verification processes. Uncertainty-aware classifiers become essential not just to improve the reliability of automated catalog generation but also to enable the discovery of rare and unseen astrophysical objects through principled OOD identification.

In the context of large-scale surveys, automated morphology classification problems contain some inherent limitations. One practical limitation is the scarcity of representative examples for rare morphological classes. As the number of classes increases, the number of available training samples for rare morphological classes decreases, which exacerbates the imbalance in the overall training and testing sets. In addition, many morphological classes are not sharply separated, and class ambiguity can complicate learning, leading to non-uniform performance gains across classes. This is particularly relevant for us, since the Galaxy Zoo DECaLS labels are derived from a fixed, extracted subset of the DECaLS archive and therefore already exhibit some degree of class imbalance.

A second limitation to be acknowledged is the increased computational cost due to MC dropout. Since uncertainty estimates are obtained from multiple stochastic forward passes, inference becomes more expensive than in a standard deterministic model. This can be managed for targeted analysis, but it can become significant when processing very large survey samples or when repeated evaluation is required.

A further limitation is interpretability. Although entropy-based and uncertainty-based metrics can identify anomalous samples, they do not directly explain why a sample is assigned high entropy or flagged as OOD. A deeper physical interpretation of such drawbacks will require future work, including inspection of individual objects and comparison with imaging artifacts, rare morphologies, and other observational effects.

These limitations do not affect the utility of the proposed framework, but they do motivate future work on class imbalance, computational efficiency, and physical interpretation of OOD scores.
We have not attempted an analysis based on morphology-aware misclassification patterns, placing particular emphasis on visually similar classes, such as ringed versus arm-dominated systems. Also, an in-depth examination of galaxies flagged as OOD will be conducted to describe whether they represent rare morphologies, artifacts, or low-quality observations.

\section*{Acknowledgments}
This work was supported by the Ministry of Education (MoE), Government of India. We appreciate the support of the IIT Hyderabad and Japan International Co-operation Agency (JICA) for access to computing facilities. {We are also grateful to the anonymous referee for constructive feedback and many useful comments on the manuscript.
\clearpage
\bibliographystyle{model2-names}
\bibliography{example}

\clearpage
\appendix
\setcounter{figure}{0}
\section{Appendix Section}
This appendix presents the supplementary qualitative visualization that complements the quantitative results reported in the main text. Grad-CAM maps provide a local visualization of the pixels seen by the classifier in different layers that contribute to decision-making. The Projection of the UMAP, on the other hand, visualizes the latent embedding learned by the shared ResNet-34 backbone and illustrates the spread of samples in the feature space under different inference strategies discussed. Conclusively, they show that the model based on IsoMaxPlus and Hybrid (IsoMaxPlus + MC dropout) produces compact, prototype-centered clusters and clearer separation between ID and OOD samples. These figures are included here to improve interpretability and to provide a comparative visualization of the proposed models.

\begin{figure}[h]
    \centering
    \includegraphics[width=0.9\columnwidth]{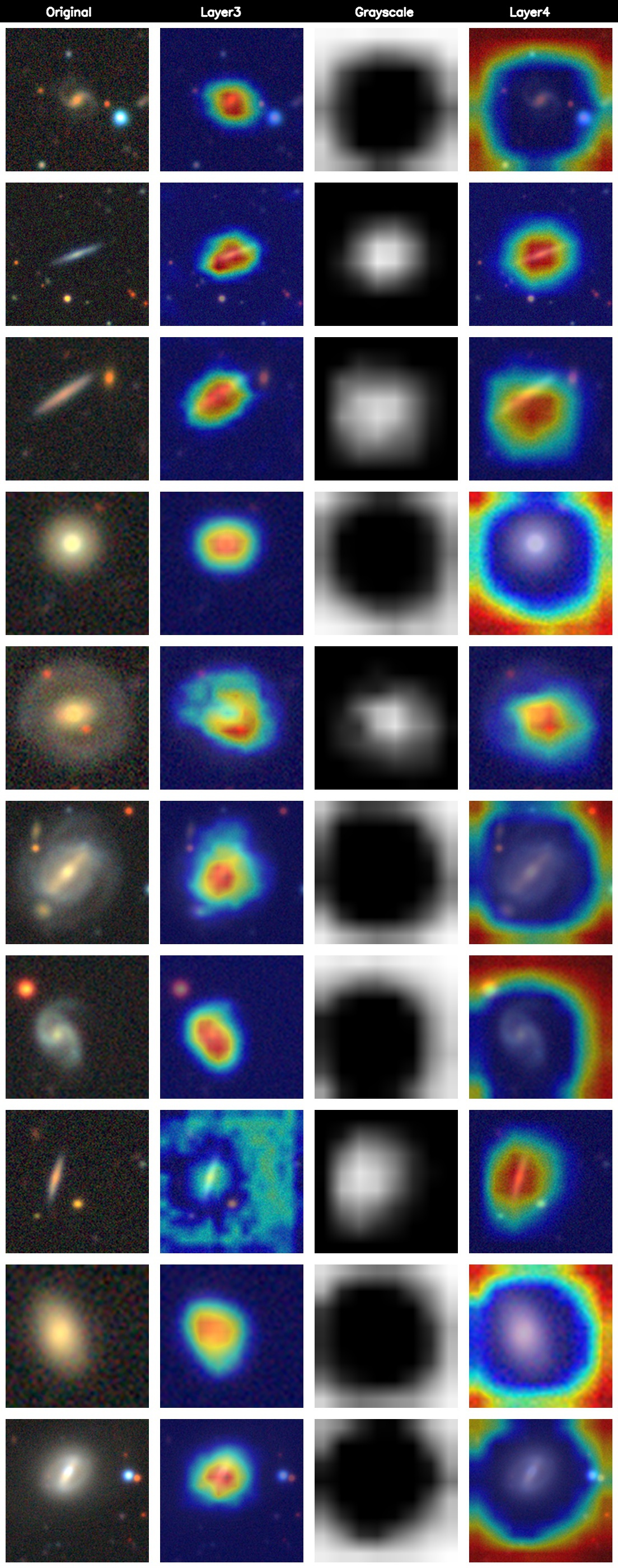}
    \caption{\textbf{Grad-CAM visualization for correctly classified galaxies.} Each row corresponds to one object, from left to right, the four-panel layout showing the original image, Layer-3 Grad-CAM, grayscale attribution, and Layer-4 Grad-CAM. The heatmaps in the correctly classified examples concentrate on physically meaningful regions. Layer-4 maps are more spatially focused than the Layer-3 maps, indicating that the deeper convolutional features align more effectively with the class-confining galaxy structure. This agreement between the attribution maps and morphology supports the interpretation that the model's correct predictions are driven by relevant features rather than by spurious backgrounds.}
    \label{fig:gradcam_correct}
\end{figure}

\begin{figure}[h]
    \centering
    \includegraphics[width=0.9\columnwidth]{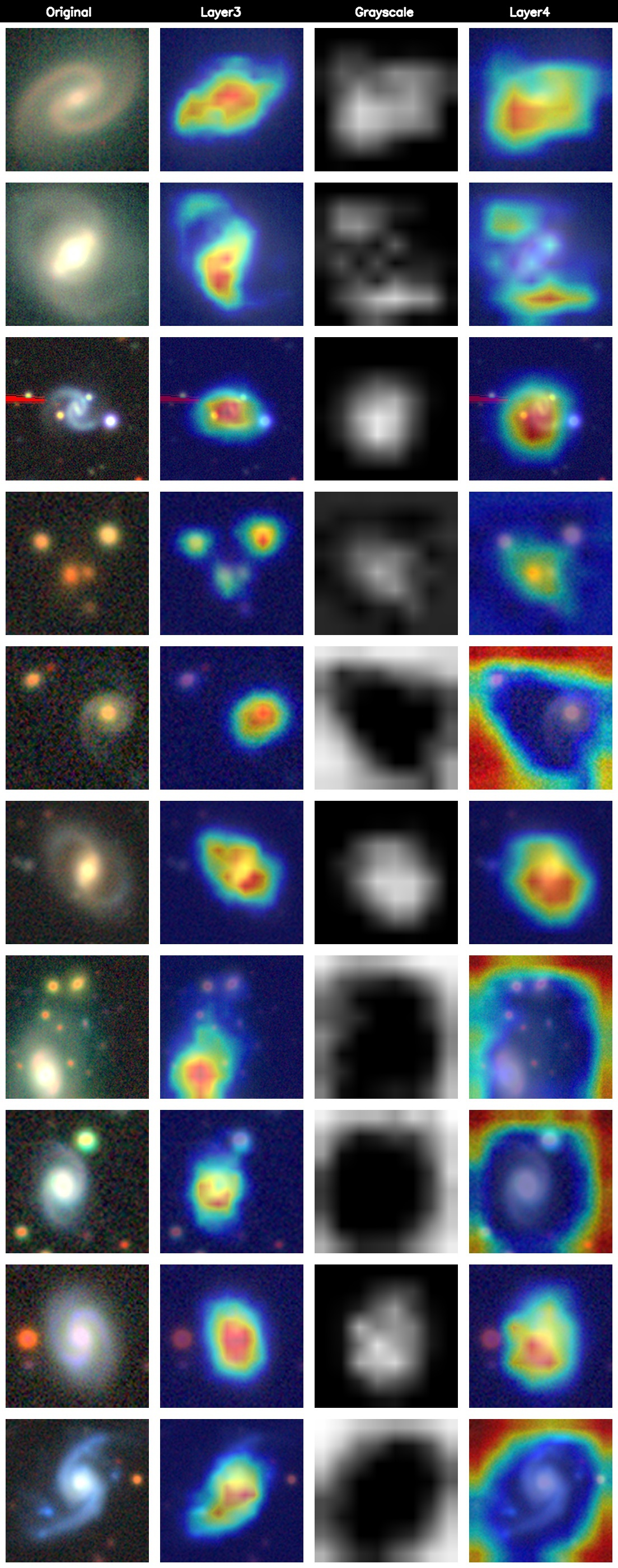}
    \caption{\textbf{Grad-CAM visualization for misclassified galaxies.} Each row shows one object; the original galaxy image in the first column to the Layer-4 attention map in the right panel. The misclassified samples show that the network often distributes attention across the central light concentration, diffuse outer structures, or nearby contaminants rather than extracting the morphology-confining regions. This pattern is consistent with ambiguous structure, projection effects, low-surface-brightness regions, or either background. It indicates that the prediction error is driven by weak or partially obscured morphology.}
    \label{fig:gradcam_wrong}
\end{figure}

\begin{figure*}[htbp]
    \centering
    \subfloat[CE baseline.]{
        \includegraphics[width=0.6\textwidth]{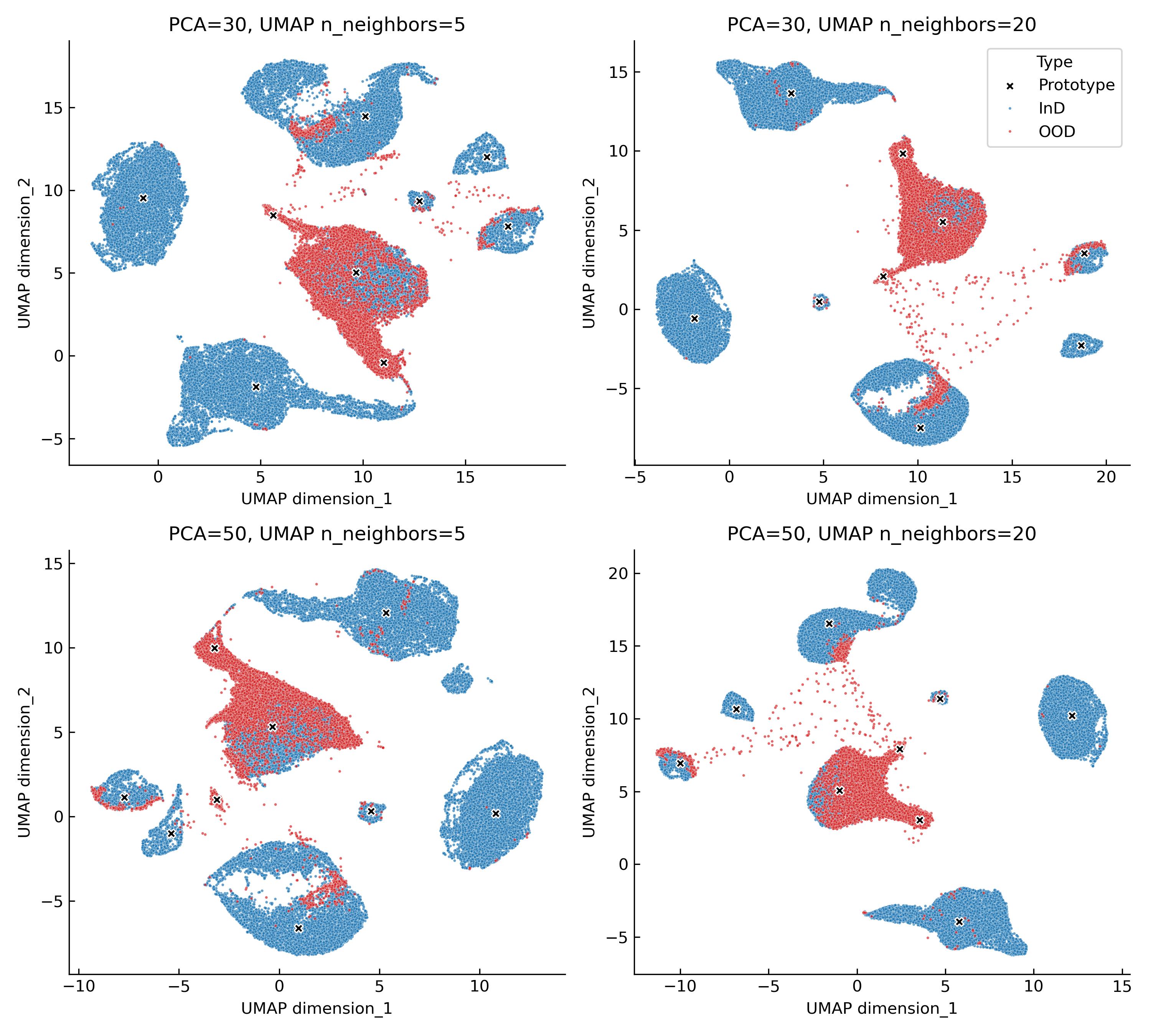}
    }
    \vspace{0.3cm}
    \subfloat[IsoMaxPlus.]{
        \includegraphics[width=0.6\textwidth]{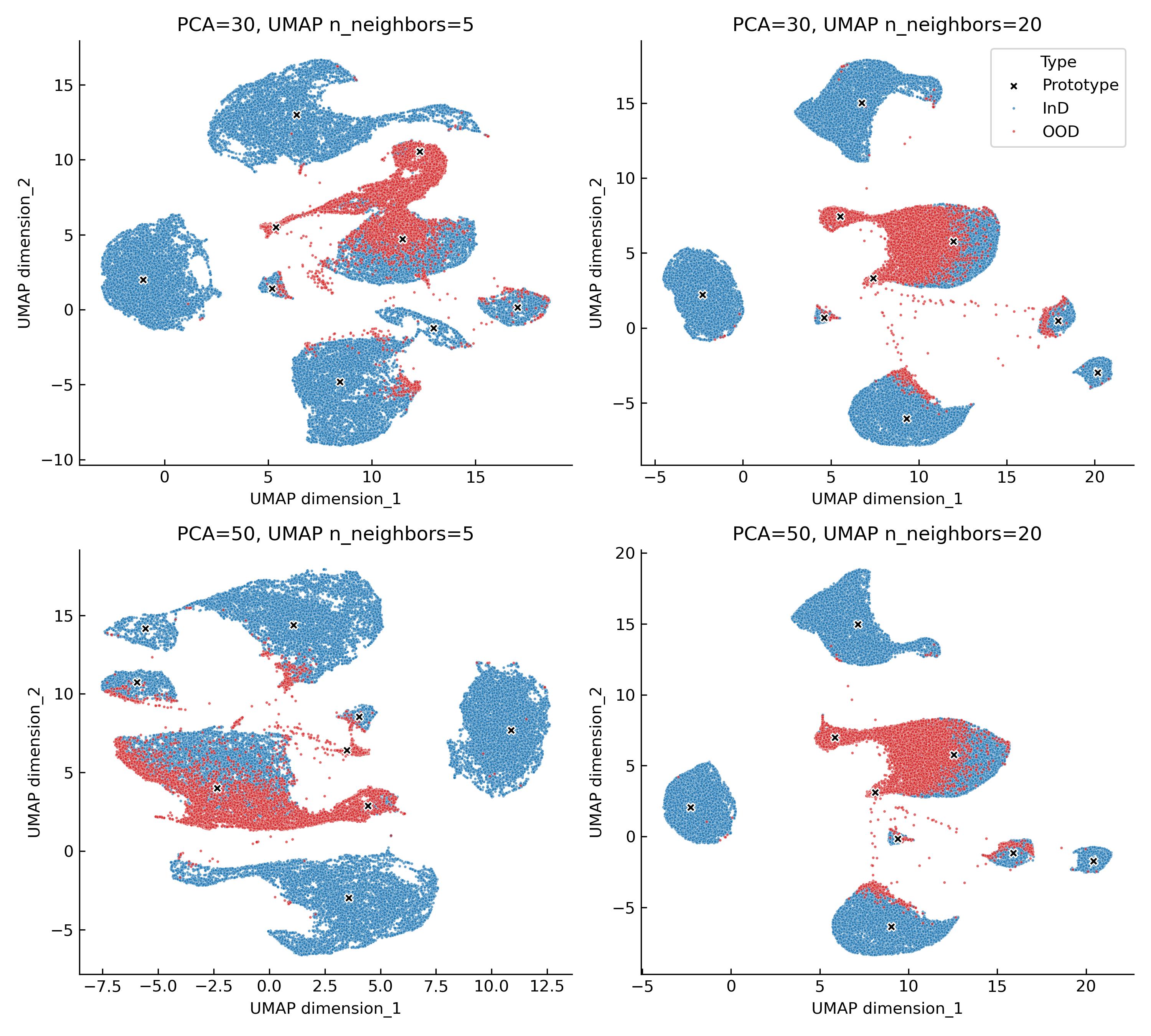}
    }
    \caption{Two-dimensional UMAP projections of the latent feature space after PCA pre-reduction to 30 and 50 components, along with UMAP neighbor size of 5 and 20. The ``x'' denotes the learned prototypes. Blue points represent the ID samples, and red dots represent OOD samples. The subplots, from top to bottom in a single column, have the same number of PCA components, with variation in UMAP neighbors; similarly, each row has similar UMAP neighbors, with variation in the number of PCA components. The various plots denotes different technique; '(a)' is for CE baseline, `(b)' is for IsoMaxPlus, when observing closely `b' shows a much compact region for OOD samples compared to `(a)' which shows the significance of respective techniques in OOD detection, some overlaps remain in middle region signifies accumulation of OOD samples near the visually similar ID class samples.}
    \label{fig:UMAP_type}
\end{figure*}

\begin{figure*}[htbp]
    \centering
    \subfloat[CE baseline.]{
        \includegraphics[width=0.6\textwidth]{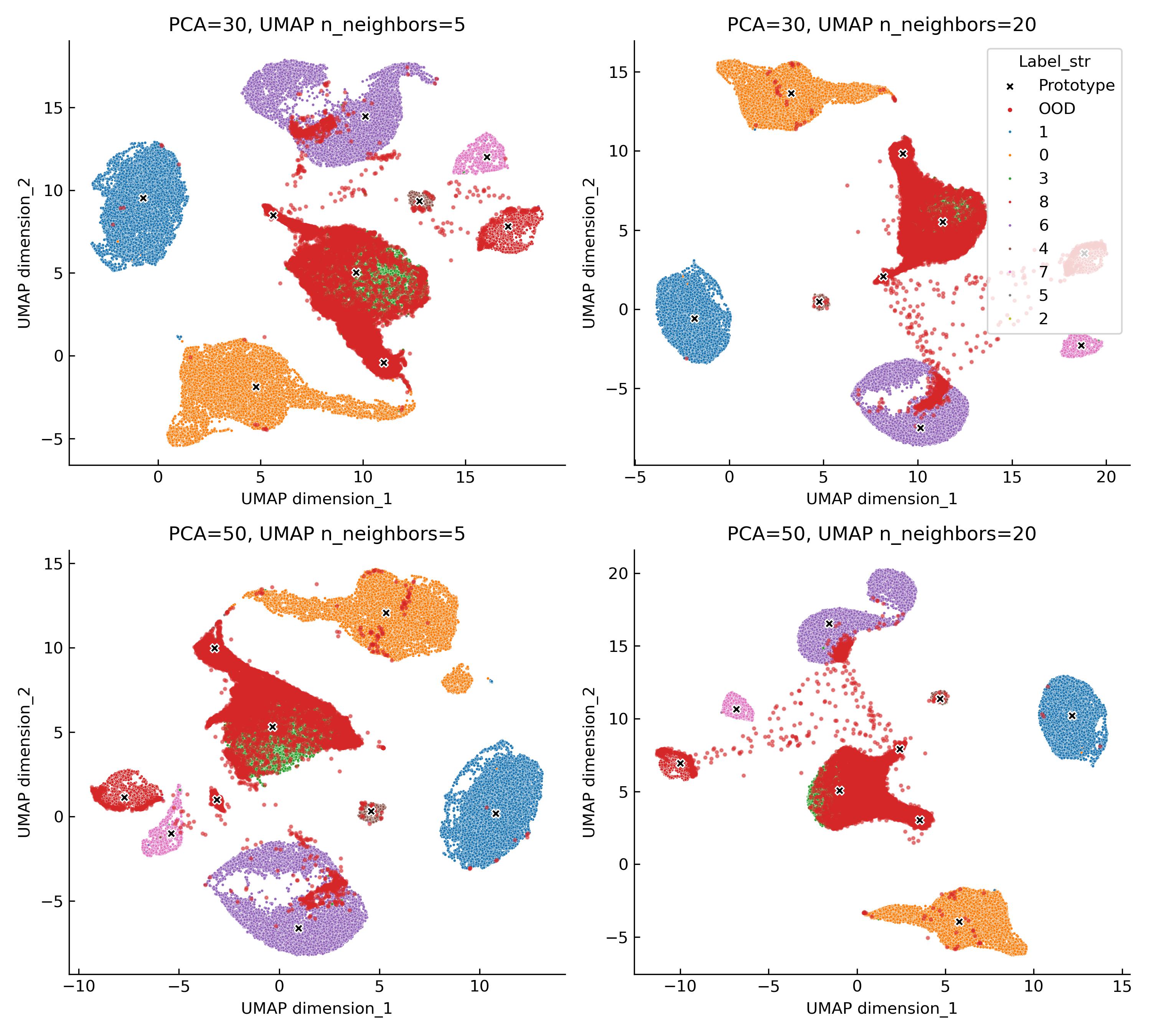}
    }
\end{figure*}
\begin{figure*}[htbp]
    \ContinuedFloat
    \centering
    \subfloat[IsoMaxPlus.]{
        \includegraphics[width=0.6\textwidth]{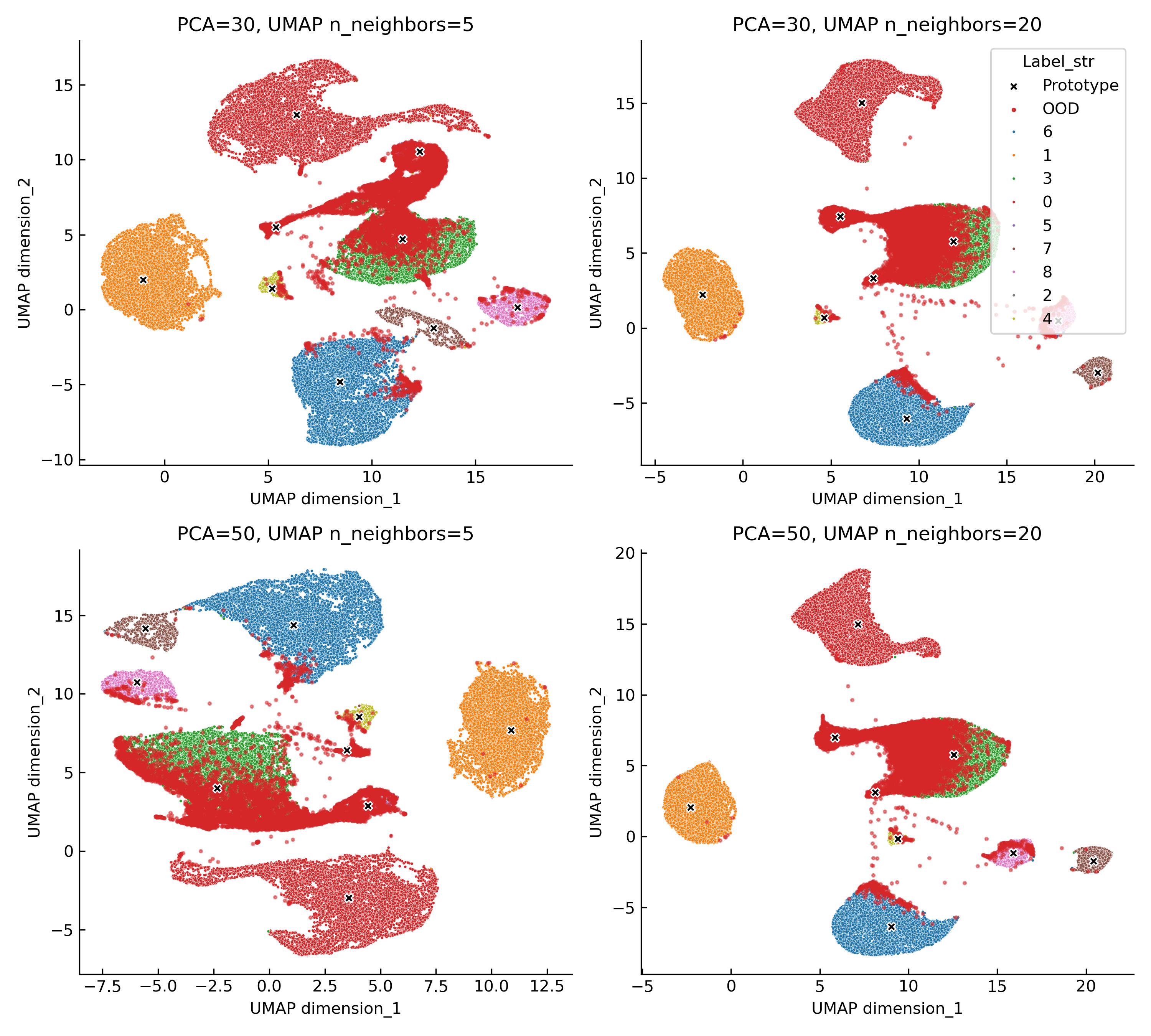}
    }
    \vspace{0.4cm}
    \caption{Two-dimensional UMAP projections of the latent feature space after PCA pre-reduction to 30 and 50 components, with 5 and 20 UMAP neighbors. The black ``x'' denotes the learned prototypes. Various colors denote different ID classes, with OOD samples represented in red. From top to bottom within a column, the number of PCA components remains the same, while from left to right within a row, the neighborhood size of UMAP remains the same. Plots `(a)' denote the CE baseline, `(b)' denotes IsoMaxPlus. These plots `b' signify the improvement in OOD detection as compared to the baseline `(a)', with a much compact spread of OOD samples in the overall Cartesian plane}
    \label{fig:UMAP_labels}
\end{figure*}

\end{document}